\documentclass[12pt]{article}
\usepackage{amsfonts,amssymb}

\catcode`@=11
\def\secteqno{\@addtoreset{equation}{section}%
\def\theequation{\thesection.\arabic{equation}}}
\catcode`@=12
\secteqno

\begin{document}
\thispagestyle{empty}

\begin{flushright}
 KEK-TH-923
\end{flushright}

\vspace{10mm}

\begin{center}
 {\Large \bf Gauge Theory on Noncommutative Supersphere}

\vspace{2mm}
 {\Large \bf from Supermatrix Model} 

\vspace{10mm}
{\large Satoshi Iso and Hiroshi Umetsu}

\vspace{5mm}

{\it 
Theory Division,\ High Energy Accelerator Research Organization (KEK),\\
Tsukuba,\ Ibaraki,\ 305-0801, Japan }

\vspace{5mm}

{\small\sf E-mails:\ satoshi.iso@kek.jp, umetsu@post.kek.jp} 
\end{center}

\vspace{10mm}

\begin{abstract}
 We construct a supermatrix model which has a classical solution
 representing the noncommutative (fuzzy) two-supersphere.
 Expanding supermatrices around the classical background, 
 we obtain a gauge theory on a noncommutative superspace on sphere.
 This theory has $osp(1|2)$ supersymmetry and $u(2L+1|2L)$ 
 gauge symmetry.
 We also discuss a commutative limit of the model keeping radius of 
 the supersphere fixed.
\end{abstract}


\newpage

\setcounter{page}{1}

\parskip 2.4mm
\section{Introduction}
Deformation of superspace by introducing noncommutativity has attracted 
much interest recently.
It is suggested that non anti-commutativity of fermionic coordinates
of superspace appears in superstring theory in the background of the RR or 
graviphoton field strength\cite{OV, deBoer, Seiberg}.  
This phenomenon is similar to the well-known case of the string theory 
in the NS-NS two form $B$ background, where the bosonic space-time
coordinates become noncommutative~\cite{Schomerus, SeibergWitten}.
The supersymmetric gauge theories and Wess-Zumino model on the
noncommutative superspace are actively studied and various aspects of 
those filed theories which include renormalizability in perturbations, 
UV/IR mixing etc. are discussed~\cite{Chepelev}-\cite{Chandrasekhar}.
There are also earlier works where noncommutative superspace was
studied~\cite{Schwarz}-\cite{Cornalba}. 

There are also analyses of noncommutative superspace by using
supermatrices~\cite{Grosse}-\cite{Shibusa}.  
Supersymmetric actions for scalar multiplets on the fuzzy
two-supersphere were constructed in \cite{Grosse} based on the 
$osp(1|2)$ graded Lie algebra. Furthermore a graded differential
calculus on the fuzzy supersphere is discussed in \cite{Grosse2}.
Supersymmetric gauge theories on this noncommutative superspace was
studied in \cite{Klimcik} by using differential forms on it.
In \cite{HIU}, noncommutative superspaces and their flat limits 
are studied by using the graded Lie algebras $osp(1|2)$, $osp(2|2)$ 
and $psu(2|2)$. 
Recently the concept of noncommutative superspace based on a
supermatrix was also introduced in proving the Dijkgraaf-Vafa
conjecture as the large N reduction~\cite{Kawai}.
Supermatrix model was also studied from the viewpoint of background 
independent formulations of matrix model which are expected to
give constructive definitions of string theories~\cite{supermatrix}.

In this paper we construct a supersymmetric gauge theory on 
the fuzzy two-supersphere based on a supermatrix model.
This is a natural extension of constructing gauge theories on 
the bosonic noncommutative space in matrix models.
In the ordinary matrix models of the IKKT type~\cite{IKKT}, 
background space-time appears as a classical background of 
matrices $A_i$ and their fluctuations around the classical solution 
are interpreted as gauge fields on this space-time.
If the classical solutions are noncommutative, we can obtain
noncommutative gauge theories~\cite{AIIKKT}.
In this approach, constructions of the open Wilson lines and background
independence of the noncommutative gauge theories become manifest
\cite{AIIKKT, IIKK, Seiberg2}.
Construction of supermatrix models whose classical backgrounds represent
a noncommutative superspace will similarly play an important role to
understand various properties of field theories on the noncommutative
superspace. 
In this paper we particularly investigate a supersymmetric gauge theory
on the fuzzy two-supersphere by using a simple supermatrix model based
on the $osp(1|2)$ graded Lie algebra. 
Noncommutative superspace coordinates $(x_i, \ \theta_\alpha)$ and gauge
superfields $(\tilde{a}_i, \ \varphi_\alpha)$ on it are combined as single
supermatrices, $A_i\sim x_i + \tilde{a}_i$ and 
$\psi_\alpha\sim \theta_\alpha+\varphi_\alpha$.
Our formulation of a supersymmetric gauge theory on the fuzzy supersphere 
has some similarities to the covariant superspace approach in the
ordinary supersymmetric gauge theories~\cite{superspace}.  
In this approach, the connection superfields on the superspace are
introduced and constraints are imposed on them to eliminate extra
degrees of freedom.
It turns out that supermatrices in our model correspond to 
the connection superfields on the noncommutative supersphere. 

This paper is organized as follows.
In section 2, we first review the construction of the fuzzy
two-supersphere based on the $osp(1|2)$ graded Lie algebra.
The representations of $osp(1|2)$ are explained and fields on the
fuzzy space are introduced as polynomials of the representation matrices 
of the $osp(1|2)$ generators. 
In section 3, we construct a supermatrix model which has a classical
solution corresponding to the fuzzy two-supersphere.
Expanding a supermatrix around this classical background we obtain 
a supersymmetric gauge theory on the fuzzy supersphere.
The action has $osp(1|2)$ supersymmetry and $u(2L+1|2L)$ gauge
symmetry. 
Then it is shown that in a commutative limit this model gives
the $U(1)$ gauge theory on a commutative sphere.
Conclusions and discussions are given in section 4. 
Brief explanations of the graded Lie algebra and supermatrix are given 
in the appendix. 

\section{Fuzzy two-supersphere}

In this section we review a construction of supermatrix models and 
field theories on the fuzzy two-supersphere based on $osp(1|2)$
algebra. This was first studied in \cite{Grosse}.
Notations and definitions used in this paper are given in the appendix. 

The graded commutation relations of $osp(1|2)$ algebra are given by 
\begin{eqnarray}
\left[\hat{l}_i, \hat{l}_j\right] &=& i\epsilon_{ijk}\hat{l}_k, \nonumber \\
\left[\hat{l}_i, \hat{v}_\alpha\right] &=&
  \frac{1}{2}\left(\sigma_i\right)_{\beta\alpha}\hat{v}_\beta, \\
\label{ospalgebra}
\left\{\hat{v}_\alpha, \hat{v}_\beta\right\} &=&
  \frac{1}{2}\left(C\sigma_i\right)_{\alpha\beta}\hat{l}_i, \nonumber
\end{eqnarray}
where $C=i\sigma_2$ is a charge conjugation matrix.
The even part of this algebra is $su(2)$ which 
is generated by $\hat{l}_i \ (i=1, 2, 3)$ and the odd generators 
$\hat{v}_\alpha \ (\alpha=1, 2)$ are $su(2)$ spinors.
Irreducible representations of $osp(1|2)$
algebra~\cite{representation} are characterized
by values of the Casimir operator 
$\hat{K}_2=\hat{l}_i\hat{l}_i
+C_{\alpha\beta}\hat{v}_\alpha \hat{v}_\beta=L(L+\frac{1}{2})$ 
where quantum number $L$ is 
called super spin and 
$L\in\mathbb{Z}_{\geq 0}/2$. 
Each representation consists of spin $L$ and $L-\frac{1}{2}$
representations of $su(2)$, $|L, l_3\rangle, |L-\frac{1}{2}, l_3\rangle$
and its dimension is $N\equiv (2L+1)+2L=4L+1$.
The explicit expressions of the generators are 
\begin{eqnarray}
 && l_i^{(L)} = \left(
	  \begin{array}{cc}
	   L_i^{(L)} & 0 \\
	   0 & L_i^{(L-1/2)}
	  \end{array}
	  \right), 
 \hspace{10mm}
 v_\alpha^{(L)} = \left(
	    \begin{array}{cc}
	     0 & V_\alpha^{(L, L-1/2)}\\
	     V_\alpha^{(L-1/2, L)} & 0
	    \end{array}
	    \right).
\end{eqnarray}
Matrix elements of $L_\pm = L_1 \pm i L_2$, $V_+=V_1$ and $V_-=V_2$ are
given by 
\begin{eqnarray}
 \langle L, l_3+1|L_+^{(L)}|L, l_3\rangle 
  &=& \sqrt{(L-l_3)(L+l_3+1)}, \nonumber \\
 \langle L, l_3-1|L_-^{(L)}|L, l_3\rangle 
  &=& \sqrt{(L+l_3)(L-l_3+1)}, \nonumber \\
 \langle L, l_3+1/2|V_+^{(L, L-1/2)}|L-1/2, l_3 \rangle 
  &=& -\frac{1}{2}\sqrt{L+l_3+\frac{1}{2}}, \nonumber \\
\langle L, l_3-1/2|V_-^{(L, L-1/2)}|L-1/2, l_3 \rangle 
  &=& -\frac{1}{2}\sqrt{L-l_3+\frac{1}{2}}, \\
\langle L-1/2, l_3+1/2|V_+^{(L-1/2, L)}|L, l_3 \rangle 
  &=& -\frac{1}{2}\sqrt{L-l_3}, \nonumber \\
\langle L-1/2, l_3-1/2|V_-^{(L-1/2, L)}|L, l_3 \rangle 
  &=& \frac{1}{2}\sqrt{L+l_3}. \nonumber 
\end{eqnarray}
These are the superstar representations of $osp(1|2)$,
\begin{eqnarray}
 {l_i^{(L)}}^\ddagger = l_i^{(L)}, \qquad 
  {v_\alpha^{(L)}}^\ddagger = -C_{\alpha\beta}v_\beta^{(L)}.
\end{eqnarray}
See the appendix for superstar conjugation $\ddagger$.

The condition $\hat{K}_2=L(L+\frac{1}{2})$ 
defines a two-dimensional supersphere.
Consider polynomials $\Phi(l_i^{(L)}, v_\alpha^{(L)})$ of 
the representation matrices $l_i^{(L)}$ and $v_\alpha^{(L)}$ 
with super spin $L$. 
Let us denote the space spanned by $\Phi(l_i^{(L)}, v_\alpha^{(L)})$ 
as ${\cal A}_L$.
The $osp(1|2)$ algebra acts on ${\cal A}_L$ by three kinds of action, 
the left action $(\hat{l}_i^L, \hat{v}_\alpha^L)$, the right action 
$(\hat{l}_i^R, \hat{v}_\alpha^R)$ and the adjoint action
$(\hat{{\cal L}}_i\equiv \hat{l}_i^L-\hat{l}_i^R, \hat{{\cal V}}_\alpha=\hat{v}_\alpha^L-\hat{v}_\alpha^R)$,
\begin{eqnarray}
 \hat{l}_i^L \Phi = l_i^{(L)} \Phi, && 
  \hat{v}_\alpha^L \Phi = v_\alpha^{(L)} \Phi, \\
 \hat{l}_i^R \Phi = \Phi l_i^{(L)}, && 
  \hat{v}_\alpha^R \Phi = \Phi  v_\alpha^{(L)}, \\
 \hat{{\cal L}}_i \Phi = [l_i^{(L)}, \Phi], && 
 \hat{{\cal V}}_\alpha \Phi = \left[v_\alpha^{(L)}, \Phi\right].
\end{eqnarray}
The right action satisfies the $osp(1|2)$ algebra with a minus sign 
$(-l_i^R, \ -v_\alpha^R)$.
The polynomials transform as $L\otimes L$ under the left and right
action of $osp(1|2)$ and can be decomposed into the irreducible
representations under the adjoint action as 
$$
L\otimes L=0\oplus \frac{1}{2} \oplus 1\oplus \cdots
\oplus 2L-\frac{1}{2}\oplus 2L.
$$
The dimension of the space spanned by these polynomials is $(4L+1)^2$.
Among them, we can define supersymmetrized matrix spherical harmonics
$Y_{km}^S(l_i^{(L)}, v_\alpha^{(L)})$ which are generalization of 
the ordinary matrix spherical harmonics to the supersphere 
(see \cite{Grosse2} for the details), 
\begin{eqnarray}
 && \left(\hat{\cal L}_i \hat{\cal L}_i 
     + C_{\alpha\beta}\hat{\cal V}_\alpha \hat{\cal V}_\beta
     \right) Y^S_{km}(l_i^{(L)}, v_\alpha^{(L)})
 = k\left(k+\frac{1}{2}\right)Y^S_{km}(l_i^{(L)}, v_\alpha^{(L)}), \\
 && \hat{\cal L}_3 Y^S_{km}(l_i^{(L)}, v_\alpha^{(L)}) 
  = m Y^S_{km}(l_i^{(L)}, v_\alpha^{(L)}).
\end{eqnarray} 
$k$ can take either an integer or a half-integer value. 
Any $N \times N$ supermatrix can be expanded in terms of
the superspherical harmonics as
\begin{eqnarray}
 \label{expansion}
  \Phi(l_i^{(L)}, v_\alpha^{(L)})
  =\sum_{k=0, 1/2, 1, \cdots}^{2L} \phi_{km}
  Y^S_{km}(l_i^{(L)}, v_\alpha^{(L)}),
\end{eqnarray} 
where the Grassmann parity of the coefficient $\phi_{km}$
is determined by the grading of the spherical harmonics.
Even (odd) spherical harmonics has nonvanishing values 
only in the diagonal (off-diagonal) blocks in its matrix form. 
We can map the supermatrix $\Phi(l_i^{(L)}, v_\alpha^{(L)})$ 
to a function on the superspace $(x_i, \theta_\alpha)$ by 
\begin{equation}
 \label{map_field}
 \Phi(l_i, v_\alpha)\longrightarrow 
  \phi(x_i, \theta_\alpha)=\sum_{k,m} \phi_{km} \ 
  y^S_{km}(x_i, \theta_\alpha),
\end{equation}
where $y^S_{km}(x_i, \theta_\alpha)$ are ordinary superspherical
functions.
A product of supermatrices is mapped to a noncommutative star product
of functions.
An explicit form of the star product is given in \cite{Balachandran}.

In addition to the $osp(1|2)$ generators $(\hat{l}_i, \hat{v}_{\alpha})$,
we can define additional generators with which they form bigger
algebra $osp(2|2)$.
These additional generators are 
\begin{eqnarray}
&& \hat{\gamma} = -\frac{1}{L+1/4}
\left(C_{\alpha \beta}\hat{v}_{\alpha} \hat{v}_{\beta} 
 + 2 L\left(L+{1 \over 2}\right)\right) \\
\label{addg}
&& \hat{d}_{\alpha} = [\hat{\gamma}, \hat{v}_{\alpha}]
=\frac{1}{2(L+1/4)}(\sigma_i)_{\beta \alpha} 
\left(\hat{v}_{\beta} \hat{l}_i + \hat{l}_i \hat{v}_{\beta}\right).
\label{addd}
\end{eqnarray}
Commutation relations for the additional generators are given by
\begin{eqnarray}
\left[\hat{\gamma}, \hat{v}_\alpha\right] &=& \hat{d}_\alpha, 
 \hspace{23mm}
 \left[\hat{\gamma}, \hat{d}_\alpha\right] = \hat{v}_\alpha,  
 \hspace{31mm}
 \left[\hat{\gamma}, \hat{l}_i\right] = 0, \nonumber \\
 \left[\hat{l}_i, \hat{d}_\alpha\right] &=&
  \frac{1}{2}\left(\sigma_i\right)_{\beta\alpha}\hat{d}_\beta,  
  \qquad
  \left\{\hat{d}_\alpha, \hat{d}_\beta\right\} =
  -\frac{1}{2}\left(C\sigma_i\right)_{\alpha\beta}\hat{l}_i, \qquad
 \left\{\hat{v}_\alpha, \hat{d}_\beta\right\} =
  -\frac{1}{4}C_{\alpha\beta}\hat{\gamma}. \nonumber 
\end{eqnarray}
The adjoint action of 
the fermionic generators $D_{\alpha}=\mbox{adj}~\hat{d}_\alpha$ 
plays a role of the covariant derivatives on the supersphere. 
On the other hand, the adjoint action 
of the original fermionic generators 
$Q_{\alpha}=\mbox{adj}~\hat{v}_\alpha$ 
are interpreted as supersymmetry generators. 
These additional generators also play an important role in 
constructing kinetic terms for a scalar multiplet on the supersphere
\cite{Grosse}.

The commutative limit is  discussed in \cite{Grosse}
and the fuzzy supersphere becomes the ordinary
two-dimensional supersphere with two real grassmannian coordinates. 
This limit can be taken by keeping the radius
of the sphere fixed and taking the large $L$ limit.

\section{Gauge theory on fuzzy supersphere}

In this section we construct a supermatrix model which has 
a classical solution representing the fuzzy supersphere.
Expanding supermatrices around the classical solution 
we obtain the action with the supersymmetry and gauge symmetry.
This is a supermatrix extension of the construction of a gauge theory on
fuzzy sphere from matrix models~\cite{IKTW}.

Let us consider a supermatrix $M$ which has the following form,
\begin{equation}
 M=A_i \otimes t_i + C_{\alpha\beta}\psi_\alpha \otimes q_\beta,
\end{equation}
where $t_i \ (i=1, 2, 3)$ and $q_\alpha \ (\alpha=1,2)$ are 
the $L=1/2$ representation matrices of the $osp(1|2)$ algebra,
\begin{eqnarray}
\label{t-q}
 && t_i=\frac{1}{2}\left(
	\begin{array}{cc}
	\sigma_i & 0 \\
	 0 & 0
	\end{array}
	\right), 
 \hspace{5mm}
q_1=\frac{1}{2}\left(
	\begin{array}{ccc}
	0 & 0 & 1 \\
	 0 & 0 & 0 \\
	 0 & 1 & 0
	\end{array}
    \right), 
\hspace{5mm}
q_2=\frac{1}{2}\left(
	\begin{array}{ccc}
	0 & 0 & 0 \\
	 0 & 0 & 1 \\
	 -1 & 0 & 0
	\end{array}
    \right).
\end{eqnarray}
$A_i$ and $\psi_\alpha$ are respectively even and odd 
$N\times N$ supermatrices with $N=4L+1$. 
We impose a reality condition $M^\ddagger = M$, 
that is $A_i^\ddagger = A_i$ and 
$\psi_\alpha^\ddagger = C_{\alpha\beta}\psi_\beta$.
We define a grading operator $B$ for $N\times N$ supermatrices
as 
\begin{eqnarray}
 B=\left(
   \begin{array}{cc}
    {\mathbf 1}_{2L+1} & 0 \\
    0 & -{\mathbf 1}_{2L}
   \end{array}\right).
\end{eqnarray}
It should be noted that $A_i$ and $\psi_\alpha$ are 
$(4L+1)\times (4L+1)$ supermatrices and can be also represented 
as polynomials of $l_i^{(L)}$ and $v_\alpha^{(L)}$ 
in a similar manner to eq.(\ref{expansion}). 
Hence they become superfields on the fuzzy supersphere 
in the commutative limit.

Let us consider the following action for $M$,
\begin{eqnarray}
  S = \frac{1}{g^2}\mbox{Str}_{(3\times 3, N\times N)}
   \left(M^3+\lambda M^2\right),
\end{eqnarray}
where $\lambda$ and $g$ are real constants.
In terms of $A_i$ and $\psi_\alpha$, it can be rewritten, by taking
traces over $(3\times 3)$ matrices, as
\begin{eqnarray}
 S = \frac{1}{g^2}\mbox{Str}_{(N\times N)}
  \left(\frac{i}{4}\epsilon_{ijk}A_iA_jA_k
   +\frac{\lambda}{2}A_iA_i 
   -\frac{3}{16}\psi_\alpha\left(\sigma_i C\right)_{\alpha\beta}
   \left[A_i, \psi_\beta\right]
   -\frac{\lambda}{2} C_{\alpha\beta}\psi_\alpha\psi_\beta\right).
\end{eqnarray} 
This action is invariant under the $osp(1|2)$ transformation
\begin{eqnarray}
 \label{osp-inv}
 \delta M=i[G, \ M], 
\end{eqnarray}
where $G$ has the form of 
\begin{eqnarray}
 \label{G}
 G=u_i {\mathbf 1}\otimes t_i + \epsilon_\alpha \otimes q_\alpha,
  \qquad G^\ddagger = G.
\end{eqnarray}
$u_i$ are Grassmann even numbers and $\epsilon_\alpha$ are defined as 
$\epsilon_\alpha=\tilde{\epsilon}_\alpha B$ where 
$\tilde{\epsilon}_\alpha$ are Grassmann odd numbers.
The parameters $u_i$ and $\tilde{\epsilon}_\alpha$ satisfy 
$(u_i)^\# =u_i$ and  
$(\tilde{\epsilon}_\alpha)^\# = C_{\alpha\beta}\tilde{\epsilon}_\beta$. 
It should be noted that $\epsilon_\alpha$ (anti-)commutes with 
(odd) even supermatrices because of the grading operator $B$ 
in $\epsilon_\alpha$.
Furthermore the action is invariant under the adjoint action of $u(2L+1|2L)$, 
\begin{eqnarray}
 \label{u-inv}
  \delta A_i=i[H, \ A_i], \quad
  \delta \psi_\alpha=i[H, \ \psi_\alpha], 
\end{eqnarray}
where $H^\ddagger = H, \  H \in u(2L+1|2L)$. 

The equations of motion are 
\begin{eqnarray}
  && i\epsilon_{ijk}A_jA_k+\frac{4\lambda}{3}A_i
  +\frac{1}{4}\left(\sigma_i C\right)_{\alpha\beta}
  \left\{\psi_\alpha, \psi_\beta\right\}=0, \\
 && \frac{3}{8}\left(\sigma_iC\right)_{\alpha\beta}
  \left[A_i,\psi_\beta\right]
  +\lambda C_{\alpha\beta}\psi_\beta=0.
\end{eqnarray} 
The model has a nontrivial classical solution representing  
the fuzzy two-supersphere
\footnote{
There are other classical solutions, e.g. trivial solution 
$A_i=\psi_\alpha =0$ and the fuzzy sphere solution 
$\displaystyle{A_i=\left(\frac{4}{3}\lambda\right) l_i^{(L)},
\psi_\alpha =0}$. 
We here concentrate on the fuzzy supersphere solution.},
\begin{eqnarray}
A_i^{cl}=\left(\frac{16}{9}\lambda\right)l_i^{(L)}, \qquad
\psi_\alpha^{cl}=\pm\left(\frac{16}{9}\lambda\right)d_\alpha^{(L)}.
\end{eqnarray}
We can choose $+$ sign in the classical solution of $\psi_\alpha$ 
without loss of generality because the action is invariant under 
$\psi_\alpha\rightarrow -\psi_\alpha$.
We note that the classical background $d_\alpha^{(L)}$ of 
$\psi_\alpha$ can be also written by $l_i^{(L)}$ and $v_\alpha^{(L)}$, 
eq.(\ref{addd}).
Expanding $A_i$ and $\psi_\alpha$ around the classical solution,
\begin{eqnarray}
 A_i=\frac{16}{9}\lambda \left(l_i^{(L)} + \tilde{a}_i\right), 
  \qquad \psi_\alpha=\frac{16}{9}\lambda 
  \left(d_\alpha^{(L)} + \varphi_\alpha\right), 
\end{eqnarray}
the action becomes
\begin{eqnarray}
 \label{action}
 S&=&\left(\frac{16}{9}\right)^2 \frac{\lambda^3}{g^2} \ 
  \mbox{Str}_{(N\times N)}
  \left\{
   \frac{2}{3}i\epsilon_{ijk}
   \left(\tilde{a}_i\left[l_j, \tilde{a}_k\right]
    +\frac{1}{3}\tilde{a}_i\left[\tilde{a}_j, \tilde{a}_k\right]
	 \right)
   +\frac{1}{2}\tilde{a}_i\tilde{a}_i \right. \nonumber \\
 && \left. +\left(\sigma_i C\right)_{\alpha\beta}
	     \left(\frac{2}{3}\tilde{a}_i
	      \left\{d_\alpha, \varphi_\beta\right\}
	      -\frac{1}{3}\varphi_\alpha
	      \left[l_i+\tilde{a}_i, \varphi_\beta\right]\right)
	     -\frac{1}{2}C_{\alpha\beta}\varphi_\alpha\varphi_\beta
    \right\} \\
 && +\frac{1}{6}\left(\frac{16}{9}\right)^2\frac{\lambda^3}{g^2}
  L\left(L+\frac{1}{2}\right). \nonumber 
\end{eqnarray}
The fluctuations $\tilde{a}_i$ and $\varphi_\alpha$ are respectively
even and odd $N\times N$ supermatrices which can be expanded 
in terms of polynomials of $l_i^{(L)}$ and $v_\alpha^{(L)}$.
Therefore they are regarded as the superfields on the fuzzy supersphere.
Although the backgrounds of $A_i$ and $\psi_\alpha$ violate the $osp(1|2)$
invariance (\ref{osp-inv}), it can be compensated 
by appropriate $u(2L+1|2L)$ transformations.
Actually the action is invariant under the following combination of 
$osp(1|2)$ and $u(2L+1|2L)$ with 
$H=u_il_i^{(L)}-\epsilon_\alpha d_\alpha^{(L)}$ 
(where $u_i$ and $\epsilon_\alpha$ are introduced in (\ref{G})), 
\begin{eqnarray}
 \delta \tilde{a}_i &=& -\epsilon_{ijk}u_j \tilde{a}_k
  + iu_j\left[l_j^{(L)}, \ \tilde{a}_i\right]
  -\frac{i}{2}(\sigma_i)_{\beta\alpha}\epsilon_\alpha\varphi_\beta
  -i\epsilon_\alpha\left[d_\alpha^{(L)}, \ \tilde{a}_i\right], \nonumber \\
 \delta\varphi_\alpha &=& 
  -\frac{i}{2}u_i(\sigma_i)_{\beta\alpha}\varphi_\beta
  +iu_i\left[l_i^{(L)}, \ \varphi_\alpha\right]
  -\frac{i}{2}(C\sigma_i)_{\alpha\beta}\epsilon_\beta\tilde{a}_i
  -i\epsilon_\beta\left\{d_\beta^{(L)} , \ \varphi_\alpha\right\}.
\end{eqnarray}
These are the supersymmetry transformations of this model.
There is also the $u(2L+1|2L)$ gauge symmetry,
\begin{eqnarray}
 \delta \tilde{a}_i &=& 
  i\left[H, \ l_i^{(L)}+\tilde{a}_i \right], \nonumber \\ 
 \delta\varphi_\alpha &=& 
  i\left[H, \ d_\alpha^{(L)} +\varphi_\alpha\right].
\end{eqnarray}
Therefore the action (\ref{action}) we obtained describes 
a supersymmetric gauge theory on the fuzzy supersphere. 

Let us consider the field theory representation of 
the supermatrix model.  
We introduce coordinates on the supersphere $(x_i, \ \theta_\alpha)$ as 
\begin{eqnarray}
 \label{map_coord}
 x_i &=& \frac{\rho}{\sqrt{L\left(L+\frac{1}{2}\right)}} \ l_i^{(L)}, \\
 \theta_\alpha &=& \frac{\rho}{\sqrt{L\left(L+\frac{1}{2}\right)}} 
  \ v_\alpha^{(L)},
\end{eqnarray}
where $\rho$ is a real constant.
These coordinates parametrize the noncommutative supersphere with radius
$\rho$: $x_ix_i+C_{\alpha\beta}\theta_\alpha \theta_\beta = \rho^2$.
The noncommutativity parameter is given by 
$\displaystyle{\frac{\sqrt{\rho}}{L}}$.
In the $L\rightarrow\infty$ limit
\footnote{We can consider other $L\rightarrow\infty$ limits.
For instance, a flat noncommutative limit with asymmetric scalings 
for $\theta_\alpha$ is studied in \cite{HIU}.}, 
$x_i$ and $\theta_\alpha$ become commutative coordinates.
The supermatrices $\tilde{a}_i$ and $\varphi_\alpha$ are mapped to 
superfields $\tilde{a}_i(x, \theta)$ and $\varphi_\alpha (x, \theta)$
respectively as in (\ref{map_field}). 
The adjoint actions of the $osp(2|2)$ generators on supermatrices 
become the actions of the following differential operators on 
superfields~\cite{Grosse},
\begin{eqnarray}
 {\rm adj}(l_i) & \longrightarrow &
  K_i = R_i + \frac{1}{2}\theta_\alpha\left(\sigma_i\right)_{\alpha\beta}
  \frac{\partial}{\partial\theta_\beta}, \nonumber \\
 {\rm adj}(v_\alpha) & \longrightarrow &
 K_\alpha^v = \frac{1}{2}x_i\left(C\sigma_i\right)_{\alpha\beta}
  \frac{\partial}{\partial\theta_\beta} 
  -\frac{1}{2}\theta_\beta\left(\sigma_i\right)_{\beta\alpha}\partial_i,
  \nonumber \\
 {\rm adj}(d_\alpha) & \longrightarrow &
 K^d_\alpha = -\frac{r}{2}\left(1+\frac{\theta^2}{r^2}\right)
  C_{\alpha\beta}\frac{\partial}{\partial\theta_\beta}
  +\frac{1}{2r}\theta_\beta\left(\sigma_i\right)_{\beta\alpha}R_i
  -\frac{1}{2r}\theta_\alpha x_i\partial_i, \\
 {\rm adj}(\gamma) & \longrightarrow &
 K^\gamma = \frac{1}{r}x_i\left(\sigma_i\right)_{\alpha\beta}\theta_\alpha
 \frac{\partial}{\partial\theta_\beta}, \nonumber
\end{eqnarray}
where $R_i=-i\epsilon_{ijk}x_j\partial_k$.
The supertrace can be replaced by the integral on the supersphere,
\begin{eqnarray}
 \label{map_str}
 {\rm Str}\longrightarrow 
  -\frac{\rho}{2\pi}\int d^3xd^2\theta 
  \delta\left(x^2+\theta^2-\rho^2\right). 
\end{eqnarray}
By using the mapping rules (\ref{map_field}), 
(\ref{map_coord})-(\ref{map_str}), we obtain the following action 
on the noncommutative supersphere, 
\begin{eqnarray}
 S&=&\left(-\frac{\rho}{2\pi}\right)
 \left(\frac{16}{9}\right)^2 \frac{\lambda^3}{g^2} \ 
  \int d^3x d^2\theta \delta(x^2+\theta^2-\rho^2)
  \left\{
   \frac{2}{3}i\epsilon_{ijk}
   \left(\tilde{a}_i K_j \tilde{a}_k
    +\frac{1}{3}\tilde{a}_i\left[\tilde{a}_j, \tilde{a}_k\right]
	 \right)
   +\frac{1}{2}\tilde{a}_i\tilde{a}_i \right. \nonumber \\
 && \left. +\left(\sigma_i C\right)_{\alpha\beta}
	     \left(\frac{2}{3}\tilde{a}_i
	      K^d_\alpha\varphi_\beta
	      -\frac{1}{3}\varphi_\alpha
	      \left(
	      K_i \varphi_\beta + [\tilde{a}_i, \varphi_\beta]
	      \right)\right)
	     -\frac{1}{2}C_{\alpha\beta}\varphi_\alpha\varphi_\beta
    \right\}_* \\
 && +\frac{1}{6}\left(\frac{16}{9}\right)^2\frac{\lambda^3}{g^2}
  L\left(L+\frac{1}{2}\right). \nonumber 
\end{eqnarray}
Here $*$ indicates the star product on the fuzzy supersphere
\cite{Balachandran}.

Next we consider a commutative limit of the model.
This limit is given by the $L \rightarrow\infty$ limit 
keeping the radius of the supersphere fixed.
The supermatrices $\tilde{a}_i$ and $\varphi_\alpha$ become 
superfields which can be expanded as follows,
\begin{eqnarray}
 \label{mapping}
 \tilde{a}_i(x, \theta) &=& a_i(x)+\xi_{i\alpha}(x)\theta_\alpha
  +\left(b_i(x)+\frac{1}{2r^2}x_j\partial_j a_i(x)\right)\theta^2, \\
 \varphi_\alpha (x, \theta) &=& \zeta_\alpha(x) 
  +(\sigma_\mu)_{\beta\alpha}c_\mu(x)\theta_\beta
  +\left(\chi_\alpha(x) 
    + \frac{1}{2r^2}x_j\partial_j \zeta_\alpha(x)\right)\theta^2,
\end{eqnarray}
where $r^2=x_ix_i$, 
$\theta^2=C_{\alpha\beta}\theta_\alpha\theta_\beta$ 
and $\mu=0, 1, 2, 3$.  
$a_i, b_i$ and $c_\mu$ are bosonic,  
$\xi_{i\alpha}, \zeta_\alpha$ and $\chi_\alpha$ are fermionic fields 
on the supersphere.
The $u(2L+1|2L)$ gauge parameter $H(l_i^{(L)}, v_\alpha^{(L)})$ becomes 
a superfield $H=h(x)+h_\alpha(x)\theta_\alpha + f(x)\theta^2$ 
where $h(x), f(x)$ are bosonic fields and $h_\alpha(x)$ are fermionic 
fields. We can fix the gauge degrees of freedom corresponding to 
$h_\alpha(x)$ and $f(x)$ by setting 
$C_{\alpha\beta}\theta_\alpha\varphi_\beta =0$ which means 
$c_0=\zeta_\alpha =0$.
In this gauge, we obtain the action in the commutative limit,
\begin{eqnarray}
 S &=& 
  \left(-\frac{\rho}{2\pi}\right)\left(\frac{16}{9}\right)^2
  \frac{\lambda^3}{g^2}
  \int d\Omega 
  \left[-\frac{i}{3\rho}\epsilon_{ijk}a_i R_j a_k
  +\frac{4i}{3}\rho\epsilon_{ijk}a_i R_j b_k 
  +\frac{i}{3}\epsilon_{ijk}a_i R_j c_k \right. \nonumber \\
 && +\frac{2}{3}\rho^2 b_ic_i
  +\frac{i}{3}\rho\epsilon_{ijk} c_i R_j c_k 
  -\frac{1}{4\rho}a_ia_i + \rho a_ib_i + \frac{2}{3}\rho c_ic_i
   \nonumber \\
 &&  -\frac{i}{3}\epsilon_{ijk}C_{\alpha\beta}
  \xi_{i\alpha}R_j \xi_{k\beta}
  +\frac{i}{6}\rho \epsilon_{ijk}(C\sigma_i)_{\alpha\beta}
  \xi_{j\alpha}\xi_{k\beta} \nonumber \\
 && \left. -\frac{1}{4}C_{\alpha\beta}\xi_{i\alpha}\xi_{i\beta}
  +\frac{1}{3}\rho^2 (\sigma_i)_{\alpha\beta}
  \xi_{i\beta}\chi_\alpha\right].
\end{eqnarray}
Here we have taken $L\rightarrow\infty$ commutative limit and 
dropped terms like $[a_i, a_j]_*$.
The auxiliary fields $b_i$ and $\chi_\alpha$ can be integrated out.
This leads to the following constraints,
\begin{eqnarray}
 c_i &=& -\frac{3}{2\rho}a_i-\frac{2i}{\rho}\epsilon_{ijk}R_j a_k, \\
 \xi_\alpha^{(\frac{1}{2})} & \equiv & 
  (\sigma_i)_{\alpha\beta}\xi_{i\beta}=0.
\end{eqnarray}
Then the action can be simplified as 
\begin{eqnarray}
  S &=& 
  \left(-\frac{\rho}{2\pi}\right)\left(\frac{16}{9}\right)^2
  \frac{\lambda^3}{g^2}
  \int d\Omega \left[
  -\frac{2}{3\rho}F_{ij}F_{ij}
  +\frac{2i}{3\rho}\epsilon_{ijk}\left(R_l F_{li}\right)F_{jk}
  -\frac{i}{12\rho}\left(\epsilon_{ijk}a_i R_j a_k -ia_ia_i\right)
	       \right. \nonumber \\
 && \left.
     -\frac{i}{3}\rho\epsilon_{ijk}
     \xi_{i\alpha}^{(\frac{3}{2})}
     \left(C_{\alpha\beta}R_j 
      - \frac{1}{2}(C\sigma_j)_{\alpha\beta}\right)
     \xi_{k\beta}^{(\frac{3}{2})}
    -\frac{1}{4}\rho C_{\alpha\beta}
    \xi_{i\alpha}^{(\frac{3}{2})}\xi_{i\beta}^{(\frac{3}{2})}\right],
\end{eqnarray}
where 
$\xi_{i\alpha}^{(\frac{3}{2})}=\xi_{i\alpha}
-\frac{1}{3}(\sigma_i)_{\alpha\beta}\xi_{\beta}^{(\frac{1}{2})}$
and $F_{ij}=R_ia_j-R_ja_i-i\epsilon_{ijk}a_k$.
This theory is invariant under the $U(1)$ gauge transformations
\begin{eqnarray}
 \delta a_i = R_i h(x), \qquad
  \delta \psi_\alpha = 0,
\end{eqnarray}  
where the gauge parameter $h(x)$ is a remnant of the $u(2L+1|2L)$ 
transformation.
The supersymmetries which are combinations of the $osp(1|2)$ and 
appropriate $u(2L+1|2L)$ transformations are not manifest because 
we have fixed the gauge degrees of freedom corresponding to 
$u(2L+1|2L)$.
The dynamical variables of the action are the gauge field 
$a_i \ (i=1, 2, 3)$ and the fermion $\xi_{i\alpha}$ with spin $\frac{3}{2}$
under $su(2)$. The normal component of $a_i$ becomes a
two-dimensional scalar on the sphere. 
Though this model has gauge symmetry and
supersymmetry, it is different from the ordinary supersymmetric gauge
theory in $D=2$ and its physical interpretation is not very clear. 

Our construction of the supersymmetric gauge theory 
is similar to the covariant superspace approach for the ordinary
supersymmetric gauge theories~\cite{superspace}.
In this approach the connections on the superspace 
which are described by superfields are introduced. 
Then the conventional constraints and the integrability conditions 
of the covariant derivatives are imposed in order to eliminate 
extra degrees of freedom.
The connections on the superspace correspond to the supermatrix 
$A_i$ and $\psi_\alpha$ in our model.
However there seems to be no appropriate condition, which preserve 
the $osp(1|2)$ symmetry, to eliminate extra fields. 
Instead of these conditions the equations of motion of the auxiliary
fields partially play a similar role in our case.

Although we here concentrated on the construction of the $U(1)$ gauge
theory on the fuzzy supersphere,
a generalization to $U(k)$ gauge theory can be easily realized by 
the following replacement,
\begin{eqnarray}
A_i \rightarrow  \sum_{a=1}^{k^2}A_i^a \otimes T^a, \qquad
\psi_\alpha \rightarrow \sum_{a=1}^{k^2}\psi_\alpha^a \otimes T^a,
\end{eqnarray}
where $T^a \ (a=1, 2, \cdots, k^2)$ are the generators of $U(k)$.

\section{Conclusions and discussions}
In this paper, we have constructed a supermatrix model which has a
classical solution representing the fuzzy two-supersphere.
We obtained a supersymmetric gauge theory on this noncommutative
superspace by expanding supermatrices around this background. 
In this formulation, the supermatrices which are the fluctuations around 
the classical background correspond to the superfields on the fuzzy
supersphere. 
This model has $osp(1|2)$ symmetry, which is the supersymmetry 
of the model, and $u(2L+1|2L)$ gauge symmetries.
The classical backgrounds corresponding to the fuzzy two-supersphere
violate the $osp(1|2)$ symmetry, but the action is still invariant 
under the $osp(1|2)$ transformations supplemented by an appropriate 
$u(2L+1|2L)$ transformation compensating the violation.
Then we took a commutative limit keeping the radius of the
supersphere fixed. The supermatrices such as the gauge fields and 
the gauge parameters become superfields on a commutative supersphere 
in this limit. 
After partially gauge fixing and integrating out some auxiliary fields 
in the superfields, we obtained a $U(1)$ gauge theory on the supersphere.
In the derived action, the supersymmetry is not manifest due to our gauge 
fixing condition. It is easy to generalize our construction 
to the $U(k) \ (k>1)$ gauge theory on the fuzzy supersphere.

The construction of the gauge theory on the fuzzy supersphere we
considered here has similarities to the covariant superspace approach
in the ordinary supersymmetric gauge theories. 
The supermatrices $A_i$ and $\psi_\alpha $ in our model correspond 
to connection superfields on noncommutative superspace.
The covariant superspace approach can be applied to supersymmetric gauge
theories in higher dimensions, e.g. $D=4, \ {\cal N}=1$ super Yang-Mills
theory. 
${\cal N}=\frac{1}{2}$ super Yang-Mills theory~\cite{OV, Seiberg} are
derived by introducing noncommutativity only between chiral fermionic
coordinates in the ${\cal N}=1$ superspace.
Although this theory is not written completely by supermatrices
because bosonic and half fermionic coordinates are still commutative, 
it can be described by an extension of the covariant superspace approach to 
the ${\cal N}=\frac{1}{2}$ noncommutative superspace. 
It is interesting to construct a supermatrix model whose classical
solution is the four-dimensional noncommutative superspace and quantum
fluctuations around it describe the super Yang-Mills theory.

It would be interesting to study the graded unitary group symmetry
$U(M|N)$ which supermatrix models possess.
In type IIB matrix model~\cite{IKKT}, $U(N)$ gauge symmetry can be
regarded as a matrix regularization of the area preserving
diffeomorphism in the Schild type action of the type IIB
Green-Schwarz string.
There will be a possibility where the graded unitary symmetry appears as
matrix regularization of a world sheet symmetry of covariant
formulations of superstring theories, e.g. superembeddings. 

\vspace{5mm}
{\bf Acknowledgments}\\
The work of H.U. is supported in part by JSPS Research Fellowships for
Young Scientists.

\appendix
\section{Notations and definitions}
In the appendix we briefly explain definitions and notations 
related to the graded Lie algebra and supermatrix.
More complete explanations can be seen e.g. in \cite{Cornwell, Frappat}.
We denote the space of Grassmann odd numbers as ${\mathbf B}$, 
a graded algebra as ${\cal G}$
and its even (odd) part as ${\cal G}_0 \ ({\cal G}_1)$. 
\begin{enumerate}
 \item Star and superstar for Grassmann number 
       \begin{eqnarray}
       \mbox{star :} && 
	(c\theta_i)^*=\bar{c}\theta_i^*, \quad 
	\theta_i^{**}=\theta_i, \quad
	(\theta_i\theta_j)^*=\theta_j^*\theta_i^*, \nonumber \\
	\mbox{superstar :} &&
	 (c\theta_i)^\#=\bar{c}\theta_i^\#, \quad 
	\theta_i^{\#\#}=-\theta_i, \quad
	(\theta_i\theta_j)^\#=\theta_i^\#\theta_j^\#, 
       \end{eqnarray}
       where $\theta_i \in {\mathbf B}$ and $c \in {\mathbb C}$.
 \item Adjoint and superadjoint for graded Lie algebra \\
       adjoint:
       \begin{eqnarray}
	&& \mbox{i.} \quad X \in {\cal G}_i \longrightarrow X^\dagger 
	 \in {\cal G}_i \quad \mbox{for \ } i=0, 1 \nonumber \\
	&& \mbox{ii.} \quad \left(aX+bY\right)^\dagger
	 =\bar{a}X^\dagger + \bar{b}Y^\dagger, \hspace{80mm} \ \\
	&& \mbox{iii.} \quad [X, \ Y\}^\dagger
	 =[Y^\dagger, \ X^\dagger\},  \nonumber \\
	&& \mbox{iv.} \quad \left(X^\dagger\right)^\dagger = X,
	 \nonumber
       \end{eqnarray}
       superadjoint:
       \begin{eqnarray}
	&& \mbox{i.} \quad X \in {\cal G}_i \longrightarrow X^\ddagger 
	 \in {\cal G}_i \quad \mbox{for \ } i=0, 1 \nonumber \\
	&& \mbox{ii.} \quad \left(aX+bY\right)^\ddagger
	 =\bar{a}X^\ddagger + \bar{b}Y^\ddagger, \hspace{80mm} \ \\
	&& \mbox{iii.} \quad [X, \ Y\}^\ddagger
	 =(-1)^{{\rm deg}X \cdot {\rm deg}Y}[Y^\ddagger, \ X^\ddagger\},  
	 \nonumber \\
	&& \mbox{iv.} \quad \left(X^\ddagger\right)^\ddagger 
	 = (-1)^{{\rm deg}X} X,
	 \nonumber
       \end{eqnarray}
       where $X, Y \in {\cal G}, \ a, b \in {\mathbb C}$. 
 \item Supermatrix \\
       $(m+n)\times (m+n)$ supermatrix $M$ has the form 
       \begin{eqnarray}
	\label{supermatrix}
	 M=\left(
	    \begin{array}{cc}
	     A & B \\
	     C & D
	    \end{array}\right),
       \end{eqnarray}
       where $A, B, C$ and $D$ are respectively
       $m \times m, m \times n, n \times m$ and $n \times n$ matrices.
       Even supermatrix (deg$M=0$) has Grassmann even components in 
       $A$ and $D$, and Grassmann odd components in $B$ and $C$.
       Odd supermatrix (deg$M=1$) has Grassmann odd components in 
       $A$ and $D$, and Grassmann even components in $B$ and $C$.
 \item Transpose and supertranspose for supermatrix \\
       transpose:
       \begin{eqnarray}
	M^t=\left(
	    \begin{array}{cc}
	     A^t & C^t \\
	     B^t & D^t
	    \end{array}\right),
       \end{eqnarray}
       where $A^t$ denotes the ordinary transpose of $A$, 
       and $(MN)^t \neq N^tM^t$.\\
       supertranspose:
       \begin{eqnarray}
	&& M^{st}=\left(
		   \begin{array}{cc}
		    A^t & (-1)^{{\rm deg}M}C^t \\
		    -(-1)^{{\rm deg}M}B^t & D^t
		   \end{array}\right), \nonumber \\
	&& \left(M^{st}\right)^{st}
	 =\left(
	  \begin{array}{cc}
	   A & -B \\
	   -C & D
	  \end{array}\right), \\
	&&\left(MN\right)^{st}=(-1)^{{\rm deg}M{\rm deg}N}N^{st}M^{st}.
	 \nonumber
       \end{eqnarray}
 \item Adjoint and superadjoint for supermatrix \\
       adjoint:
       \begin{eqnarray}
	&& M^\dagger = \left(M^t\right)^*, \nonumber \\
	&& \left(MN\right)^\dagger = N^\dagger M^\dagger, \\
	&& \left(M^\dagger\right)^\dagger = M. \nonumber
       \end{eqnarray}
       superadjoint:
       \begin{eqnarray}
	&& M^\ddagger = \left(M^{st}\right)^\#, \nonumber \\
	&& \left(MN\right)^\ddagger 
	 = (-1)^{{\rm deg}M{\rm deg}N}N^\ddagger M^\ddagger, \\
	&& \left(M^\ddagger\right)^\ddagger 
	 = (-1)^{{\rm deg}M}M. \nonumber
       \end{eqnarray}
 \item Supertrace
       \begin{eqnarray}
	&& {\rm Str} M = {\rm tr} A - (-1)^{{\rm deg}M} {\rm tr}D,
	 \nonumber \\
	&& {\rm Str} (M^{st})={\rm Str} M, \\
	&& {\rm Str}(MN)=(-1)^{{\rm deg}M{\rm deg}N}{\rm Str}(NM)
	 \nonumber 
       \end{eqnarray}
       where $M$ has the form (\ref{supermatrix})
 \item Scalar multiplication of a supermatrix by a Grassmann number
       \begin{eqnarray}
	bM &=& 
	 \left(
	  \begin{array}{cc}
	   b{\mathbf 1} & 0 \\
	   0 & (-1)^{{\rm deg}b}b{\mathbf 1}
	  \end{array}
	 \right)
	 \left(
	  \begin{array}{cc}
	   A & B \\
	   C & D
	  \end{array}
	 \right) \\
	Mb &=& \left(
	  \begin{array}{cc}
	   A & B \\
	   C & D
	  \end{array}
	 \right)
	\left(
	 \begin{array}{cc}
	  b{\mathbf 1} & 0 \\
	  0 & (-1)^{{\rm deg}b}b{\mathbf 1}
	 \end{array}
	\right)
       \end{eqnarray}
       where $b$ is a Grassmann number and $M$ is a supermatrix 
       (\ref{supermatrix}).
\end{enumerate}



\begin{thebibliography}{99}
\bibitem{OV} 
	H.~Ooguri and C.~Vafa,
	arXiv:hep-th/0302109, \ 
	hep-th/0303063.
 \bibitem{deBoer}
	J.~de Boer, P.~A.~Grassi and P.~van Nieuwenhuizen,
	arXiv:hep-th/0302078.
 \bibitem{Seiberg} 
	N.~Seiberg,
	JHEP {\bf 0306}, 010 (2003)
	[arXiv:hep-th/0305248].
 \bibitem{Schomerus}
	V.~Schomerus,
	JHEP {\bf 9906} (1999) 030
	[arXiv:hep-th/9903205].
 \bibitem{SeibergWitten}
	N.~Seiberg and E.~Witten,
	JHEP {\bf 9909} (1999) 032
	[arXiv:hep-th/9908142].
 \bibitem{Chepelev}
	I.~Chepelev and C.~Ciocarlie,
	JHEP {\bf 0306} (2003) 031
	[arXiv:hep-th/0304118].
 \bibitem{Britto:2003aj}
	R.~Britto, B.~Feng and S.~J.~Rey,
	JHEP {\bf 0307} (2003) 067
	[arXiv:hep-th/0306215].
 \bibitem{Berkovits:2003kj}
	N.~Berkovits and N.~Seiberg,
	JHEP {\bf 0307} (2003) 010
	[arXiv:hep-th/0306226].
 \bibitem{Terashima:2003ri}
	S.~Terashima and J.~T.~Yee,
	arXiv:hep-th/0306237.
 \bibitem{Ferrara:2003xy}
	S.~Ferrara, M.~A.~Lledo and O.~Macia,
	JHEP {\bf 0309} (2003) 068
	[arXiv:hep-th/0307039].
 \bibitem{Araki:2003se}
	T.~Araki, K.~Ito and A.~Ohtsuka,
	arXiv:hep-th/0307076.
 \bibitem{Britto:2003ak}
	R.~Britto, B.~Feng and S.~J.~Rey,
	JHEP {\bf 0308} (2003) 001
	[arXiv:hep-th/0307091].
 \bibitem{Grisaru:2003fd}
	M.~T.~Grisaru, S.~Penati and A.~Romagnoni,
	JHEP {\bf 0308} (2003) 003
	[arXiv:hep-th/0307099].
 \bibitem{Britto:2003kg}
	R.~Britto and B.~Feng,
	arXiv:hep-th/0307165.
 \bibitem{Romagnoni:2003xt}
	A.~Romagnoni,
	JHEP {\bf 0310} (2003) 016
	[arXiv:hep-th/0307209].
 \bibitem{Chaichian:2003dp}
	M.~Chaichian and A.~Kobakhidze,
	arXiv:hep-th/0307243.
 \bibitem{Lunin:2003bm}
	O.~Lunin and S.~J.~Rey,
	JHEP {\bf 0309} (2003) 045
	[arXiv:hep-th/0307275].
 \bibitem{Ivanov:2003te}
	E.~Ivanov, O.~Lechtenfeld and B.~Zupnik,
	arXiv:hep-th/0308012.
 \bibitem{Ferrara:2003xk}
	S.~Ferrara and E.~Sokatchev,
	arXiv:hep-th/0308021.
 \bibitem{Berenstein:2003sr}
	D.~Berenstein and S.~J.~Rey,
	arXiv:hep-th/0308049.
 \bibitem{Abbaspur:2003ss}
	R.~Abbaspur,
	arXiv:hep-th/0308050.
 \bibitem{Bars:2003dq}
	I.~Bars, C.~Deliduman, A.~Pasqua and B.~Zumino,
	arXiv:hep-th/0308107.
 \bibitem{Imaanpur:2003jj}
	A.~Imaanpur,
	JHEP {\bf 0309} (2003) 077
	[arXiv:hep-th/0308171].
 \bibitem{Alishahiha:2003kg}
	M.~Alishahiha, A.~Ghodsi and N.~Sadooghi,
	arXiv:hep-th/0309037.
 \bibitem{Mikulovic:2003sq}
	D.~Mikulovic,
	arXiv:hep-th/0310065.
 \bibitem{Sako:2003jx}
	A.~Sako and T.~Suzuki,
	arXiv:hep-th/0309076.
 \bibitem{Chandrasekhar}
	B.~Chandrasekhar and A.~Kumar,
	arXiv:hep-th/0310137.
 \bibitem{Schwarz}
	J.~H.~Schwarz and P.~Van Nieuwenhuizen,
	Lett.\ Nuovo Cim.\  {\bf 34} (1982) 21.
 \bibitem{Ferrara:2000mm}
	S.~Ferrara and M.~A.~Lledo,
	JHEP {\bf 0005} (2000) 008
	[arXiv:hep-th/0002084].
 \bibitem{Kosinski:2000xu}
	P.~Kosinski, J.~Lukierski and P.~Maslanka,
	arXiv:hep-th/0011053.
 \bibitem{Klemm}
	D.~Klemm, S.~Penati and L.~Tamassia,
	Class.\ Quant.\ Grav.\  {\bf 20} (2003) 2905
	[arXiv:hep-th/0104190].
 \bibitem{Cornalba}
	L.~Cornalba, M.~S.~Costa and R.~Schiappa,
	arXiv:hep-th/0209164.
 \bibitem{Grosse}
	 H.~Grosse, C.~Klimcik and P.~Presnajder,
	 Commun.\ Math.\ Phys.\  {\bf 185}, 155 (1997)
	 [arXiv:hep-th/9507074].
 \bibitem{Grosse2}
	 H.~Grosse and G.~Reiter,
	 Jour.\ Geom.\ Phys.\ {\bf 28}, 349 (1998)
 \bibitem{Klimcik}
	C.~Klimcik,
	Commun.\ Math.\ Phys.\  {\bf 206} 567 (1999)
	[arXiv:hep-th/9903112].
 \bibitem{HIU}
	M.~Hatsuda, S.~Iso and H.~Umetsu,
	Nucl.\ Phys.\ B {\bf 671} 217 (2003) 
	[arXiv:hep-th/0306251].
 \bibitem{Park:2003ku}
	J.~H.~Park,
	JHEP {\bf 0309} (2003) 046
	[arXiv:hep-th/0307060].
 \bibitem{Shibusa}
	Y.~Shibusa and T.~Tada,
	arXiv:hep-th/0307236.
\bibitem{Kawai} H.~Kawai, T.~Kuroki and T.~Morita,
	Nucl.\ Phys.\ B {\bf 664} 185 (2003) 
	[arXiv:hep-th/0303210].
 \bibitem{supermatrix}
	L.~Smolin,
	arXiv:hep-th/0006137.
	T.~Azuma, S.~Iso, H.~Kawai and Y.~Ohwashi,
	Nucl.\ Phys.\ B {\bf 610}, 251 (2001)
	[arXiv:hep-th/0102168].
 \bibitem{IKKT} 
	N.~Ishibashi, H.~Kawai, Y.~Kitazawa and A.~Tsuchiya,
	Nucl.\ Phys.\ B {\bf 498}, 467 (1997)
	[arXiv:hep-th/9612115].
 \bibitem{AIIKKT} 
	H.~Aoki, N.~Ishibashi, S.~Iso, H.~Kawai, Y.~Kitazawa and T.~Tada,
	Nucl.\ Phys.\ B {\bf 565}, 176 (2000)
	[arXiv:hep-th/9908141].
 \bibitem{IIKK} N.~Ishibashi, S.~Iso, H.~Kawai and Y.~Kitazawa,
	Nucl.\ Phys.\ B {\bf 573}, 573 (2000)
	[arXiv:hep-th/9910004].
 \bibitem{Seiberg2} N.~Seiberg,
	JHEP {\bf 0009}, 003 (2000)
	[arXiv:hep-th/0008013].
 \bibitem{superspace}
	For example, 
	S.~J.~Gates, M.~T.~Grisaru, M.~Rocek and W.~Siegel,
	``Superspace, Or One Thousand And One Lessons In Supersymmetry,''
	Front.\ Phys.\  {\bf 58} (1983) 1
	[arXiv:hep-th/0108200]. \\
	M.~F.~Sohnius,
	Phys.\ Rept.\  {\bf 128} (1985) 39. \\
	J.~Wess and J.~Bagger,
	``Supersymmetry And Supergravity,''
 \bibitem{representation} 
	A.~Pais and V.~Rittenberg,
	J.\ Math.\ Phys.\  {\bf 16}, 2062 (1975)
	[Erratum-ibid.\  {\bf 17}, 598 (1976)].
	M.~Scheunert, W.~Nahm and V.~Rittenberg,
	J.\ Math.\ Phys.\  {\bf 18}, 155 (1977).
	M.~Marcu,
	J.\ Math.\ Phys.\  {\bf 21}, 1277 (1980).
 \bibitem{Balachandran}
	A.~P.~Balachandran, S.~Kurkcuoglu and E.~Rojas,
	JHEP {\bf 0207}, 056 (2002)
	[arXiv:hep-th/0204170].
 \bibitem{IKTW} S.~Iso, Y.~Kimura, K.~Tanaka and K.~Wakatsuki,
	Nucl.\ Phys.\ B {\bf 604}, 121 (2001)
	[arXiv:hep-th/0101102].
 \bibitem{Cornwell}
	 J.~F.~Cornwell,
	 ``Group Theory In Physics. Vol. 3: Supersymmetries And Infinite Dimensional Algebras,''
 \bibitem{Frappat}
	 L.~Frappat, P.~Sorba and A.~Sciarrino,
	 arXiv:hep-th/9607161.

\end{thebibliography}
\end{document}